\documentclass{elsarticle}

\biboptions{longnamesfirst,semicolon}

\title{A bounded partition approach to identifying one fake coin and its type}

\author[1]{Takehiro Tokuda}
\ead{tokuda@cs.titech.ac.jp}

\author[2]{Yoshimichi Watanabe\corref{cor1}}
\ead{nabe@yamanashi.ac.jp}

\cortext[cor1]{Corresponding author\\
E-mail addresses: tokuda@cs.titech.ac.jp (T. Tokuda),
nabe@yamanashi.ac.jp (Y. Watanabe).}

\affiliation[1]{
     organization={Emeritus Professor, Tokyo Institute of Technology},
     addressline={2-12-1, Ookayama, Meguro},
     postcode={152-8552},
     city={Tokyo},
     country={Japan}
}

\affiliation[2]{
    organization={University of Yamanashi},
    addressline={4-3-11, Takeda},
    city={Kofu},
    postcode={400-8511},
    country={Japan}
}

\usepackage{amsthm}
\newtheorem*{exam*}{Example}

\begin{document}
\begin{abstract}
Fake coin problems using balance scales to identify one fake coin and its type among $n$ coins ($n>2$) were solved by Dyson in 1946.  Dyson gave adaptive solutions with the minimum number of weighings where later weighings may be dependent on results of past weighings.  In 2003 Born et al. gave non-adaptive solutions where all weighings are predetermined.  Both solutions require the computation of a Dyson set, which is a list of placement of each coin for each weighing.  The computation of a Dyson set  requires substantial amount of time when $n$ gets larger.

We present a bounded partition approach to the fake coin problems without computing any Dyson set.  Our approach uses bounded partition of coins recursively until the problem size may be small enough to use at most one weighing.
\end{abstract}
\newtheorem{thm}{Theorem}
\newtheorem{lem}[thm]{Lemma}
\newdefinition{rmk}{Remark}

\maketitle

\section{Introduction}
We have $n$ coins ($n>2$) and they are identical in appearance. Exactly one of $n$ coins is a fake coin which is heavier or lighter than genuine coins.  All genuine coins have the same weight. The fake coin problem of $n$ coins is to identify one fake coin and its type, heavy or light, using the minimum number of weighings on balance scales.

In 1946, Dyson showed that if $6\le 2n \le 3^k-3~(k \ge 2)$, then we can solve the fake coin problems of $n$ coins using at most $k$ weighings, and if $2n > 3^k-3$, then we cannot solve the problems of $n$ coins using at most $k$ weighings.  Dyson gave adaptive solutions to the fake coin problems where later weighings may be dependent on results of past weighings \cite{Dyson}.  In 2003, Born et al. gave non-adaptive solutions to the fake coin problems where all weighings are predetermined \cite{Born}.

Both solutions use a Dyson set, which is a list of placement of each coin for each weighing. The computation of a Dyson set requires substantial amount of time when $n$ gets larger.

In this note, we present a bounded partition approach for solving the fake coin problems without computing any Dyson set.  We directly solve the fake con problems based on the size of a problem. Our approach uses bounded partition of coins recursively until the problem size may be small enough to use at most one weighing. Historically Dyson used the size of a problem for deriving
the necessary condition of solutions for the fake coin problems.\par

\section{Basic Idea}
We consider three versions of fake coin problems P[$n$], P[($m$)], and P[($x,y$)], which are closely related to solve the original problem. Notice that each weighing has the same number of coins on the left side and the right side of balance scales.\par\vspace*{5pt}

\hspace*{-15pt}
\begin{tabular}{lp{0.8\textwidth}}
P[$n$] & Original fake coin problem of $n$ coins.\\
P[($m$)] & Fake coin problem of $m$ coins where one genuine coin outside of given $m$ coins can be used to place the same number of coins on both sides of balance scales, when the number of the left side coins and the number of the right side coins differ by $1$.\\
P[($x,y$)] & Fake coin problem of $x$ heavy coin candidates and $y$ light coin candidates. A heavy coin candidate is a genuine or heavy coin, and a light coin candidate is a genuine or light coin. Also one genuine coin outside of given $x+y$ coins can be used to place the same number of coins on both sides, when the number of the left side coins and the number of the right side coins differ by $1$.\\
\end{tabular}\par\vspace*{5pt}
Notice that original problem P[$1$] has no solution, because we cannot identify the type of the given fake coin, and P[$2$] has no solution, because we cannot tell whether a heavier coin or a lighter coin is fake. Solutions of original problems P[$3$] and P[$4$] require $2$ weighings and $3$ weighings respectively.
Also P[$1$] cannot be considered as a subproblem of P[$3$] in any sense.
But if we were allowed to use one genuine coin other than given coins, then P[$(1)$] has a solution where we compare the given coin with one genuine coin once, and P[$(2)$] has a solution where we compare one of given coins with one genuine coin at most twice. Both P[$(3)$] and P[$(4)$] can be solved using 2 weighings. Naturally P[$(1)$] is a subproblem of P[$(3)$] as we observe later.

Relations of three versions of fake coin problems are as follows.
For P[$n$], if we have a partition $(x, y, m)$ where $n=x+y+m$,
$x$ coins on the left side, $y$ coins on the right side, and $m$ coins on the outside, 
then the result of one weighing makes P[$n$] into one of the following three subproblems as shown in the following description of a basic method for solving P[$n$].

\vspace*{5pt}
P[$n$]:\par
$n\rightarrow$ partition$(x,y,m)$;

If the left side goes down, P[$(x,y)$];  $\lbrace m$ coins on the outside are genuine$\rbrace$

If the right side goes down, P[$(y,x)$];  $\lbrace m$ coins on the outside are genuine$\rbrace$

If balanced, P$[(m)$]; $\lbrace x+y$ coins on both sides are genuine$\rbrace$

\vspace*{5pt}
\noindent
We respectively refer to P[$(x,y)$], P[$(y,x)$], and P[$(m)$] as the left subproblem, the right subproblem, and the balanced subproblem of P[$n$]. Notice that one genuine coin is available after the first weighing except when $m=0$. As we observe later, when the number of the outside coins is $0$ in the first weighing, one genuine coin is
available after the second weighing.

For P[$(x,y)$], if we have a partition $((x_1,y_1), (x_2,y_2), (x_3,y_3))$, where $x=x_1+x_2+x_3$, $y=y_1+y_2+y_3$, $x_1$ heavy coin candidates and $y_1$ light coin candidates on the left side, $x_2$ heavy coin candidates and $y_2$ light coin candidates on the right side, and $x_3$ heavy coin candidates and $y_3$ light coin candidates on the outside, then the result of one weighing makes P[$(x,y)$] into one of the following three subproblems as shown in the following description of a basic method for solving P[$(x, y)$]. 

\vspace*{5pt}
P[$(x,y)$]: \par 
$(x,y) \rightarrow$ partition$((x_1,y_1), (x_2,y_2), (x_3,y_3))$;

If the left side goes down, P[$(x_1,y_2)$];  $\lbrace x_2+y_1+x_3+y_3$ coins are genuine$\rbrace$

If the right side goes down, P[$(x_2,y_1)$];  $\lbrace x_1+y_2+x_3+y_3$ coins are genuine$\rbrace$

If balanced, P[$(x_3,y_3)$];  $\lbrace x_1+y_1+x_2+y_2$ coins are genuine$\rbrace$

\vspace*{5pt}
\noindent
We respectively refer to P[$(x_1,y_2)$], P[$(x_2,y_1)$], and P[$(x_3,y_3)$] as the left subproblem, the right subproblem, and the balanced subproblem of P[$(x,y)$].

For P[$(m)$], if we have a partition $(x, y, t)$ where $m=x+y+t$, then
the result of one weighing makes P[$(m)$] into one of the following three subproblems as shown in the following description of a basic method for solving P[$(m)$]. 

\vspace*{5pt}
P[$(m)$]:\par
$m \rightarrow$ partition$(x,y,t)$;

If the left side goes down, P[$(x,y)$];  $\lbrace t$ coins on the outside are genuine$\rbrace$

If the right side goes down, P[$(y,x)$]; $\lbrace t$ coins on the outside are genuine$\rbrace$

If balanced, P[(t)]; $\lbrace x+y$ coins on both sides are genuine$\rbrace$

\vspace*{5pt}
\noindent
We respectively refer to P[$(x,y)$], P[$(y,x)$], and P[$(t)$] as the left subproblem, the right subproblem, and the balanced subproblem of P[$(m)$].

For selecting an appropriate partition, we use the size of a problem. The size of a problem P, denoted by Size[ P ],  is the sum of the number of heavy coin candidates and the number of light coin candidates of P.  Initially any one of $n$ coins can be a fake coin.  Hence all $n$ coins are heavy coin candidates and all $n$ coins are light coin candidates. Therefore Size[ P[$n$] ]$=2n$. Values of the size of problems P[$n$], P[$(x, y)$], and P[$(m)$] are as follows.

 Size[ P[$n$] ]$=n+n=2n$.

 Size[ P[$(x,y)$] ]$=x+y$.

 Size[ P[$(m)$] ]$=m+m=2m$.

\noindent
After each weighing, we have a new total number of heavy coin candidates and light coin candidates. 

We use a bounded partition of coins recursively. Partition must be made to
ensure that subproblems of smaller size can be solved using fewer weighings than the given problem.
Notice that problems P[$(x,y)$] and P[$(m)$] of size $\le 3$ can be solved using at most one weighing as described in Induction Basis.

A partition of a problem P with bound $c$ is a partition where
P1 is the left subproblem of P, P2 is the right subproblem of P,
P3 is the balanced subproblem of P,
Size[ P1 ] $\le c$, Size[ P2 ] $\le c$, and Size[ P3 ] $\le c$.

Using appropriate partition bounds, our method has the following Properties \ref{prop11}, \ref{prop12}, and \ref{prop13}.
For convenience we assume problems P[$(0)$] and P[$(0,0)$] are solvable without doing anything.

\setcounter{section}{1}
\newtheorem{prop}{Property}[section]
\begin{prop} [P\lbrack n\rbrack]\label{prop11}
{\rm
If $6 \le 2n \le 3^k-3$ for $k \ge 2$, then P[$n$] is solvable with at most $k$ weighings.
}
\end{prop}

\begin{prop}[P\lbrack (m)\rbrack]\label{prop12}
{\rm
If $2m \le 3^k-1 < 3^k$ for $k \ge 1$, then P[($m$)] is solvable with at most $k$ weighings.
}
\end{prop}

\begin{prop}[P\lbrack (x,y)\rbrack]\label{prop13}
{\rm
If $x+y \le 3^k$ for $k \ge 1$, then P[($x,y$)]  is solvable with at most $k$ weighings.
}
\end{prop}

A partition of a fake coin problem P into three subproblems is well-behaved, if the following properties hold.
\begin{enumerate}[(1)]
\item Size[ P ] $\le 3^k$ for $k \ge 2$ implies that 
Size[ left subproblem of P ] $\le 3^{k-1}$, Size[ right subproblem of P ] $\le 3^{k-1}$, and
Size[ balanced subproblem of P ] $\le 3^{k-1}$.
\item Subproblems of size $\le 3$ can be solved with at most one weighing.
\end{enumerate}

To show Properties \ref{prop11}, \ref{prop12}, and \ref{prop13}, it suffices to prove that our method has the following Properties \ref{prop21}, \ref{prop22}, \ref{prop23}, and \ref{prop24} using $3^{k-1}$ as the partition bound of the three subproblems for a given problem
of size $S$ such that $3^{k-1} <  S  \le 3^k$ in P[$(m)$] and P[$(x,y)$], and
for a given problem of size $S$ such that $3^{k-1} <  S+3 \le 3^k$ in P[$n$].
Proofs of Properties \ref{prop21}, \ref{prop22}, and \ref{prop23}. are given in the Appendix.  A proof of Property \ref{prop24} is given in Induction Basis.

\setcounter{section}{2}
\setcounter{prop}{0}
\begin{prop} [Size\lbrack~P\lbrack $n$\rbrack~\rbrack ]\label{prop21}
{\rm
If $6\le$ Size[ P[$n$] ] $\le 3^k-3$ for $k \ge 2$, then 
Size[ left subproblem of P[$n$] ] $\le 3^{k-1}$, Size[ right subproblem of P[$n$] ] $\le 3^{k-1}$, and Size[ balanced subproblem of P[$n$] ] $\le 3^{k-1}$.
}
\end{prop}

\begin{prop} [Size\lbrack~P\lbrack $(m)$\rbrack~\rbrack ]\label{prop22}
{\rm
If Size[ P[$(m)$] ] $\le 3^k$ for $k \ge 2$, then 
Size[ left subproblem of P[$(m)$] ] $\le 3^{k-1}$, Size[ right subproblem of P[$(m)$] ] $\le 3^{k-1}$, and Size[ balanced subproblem of P[$(m)$] ] $\le 3^{k-1}$.
}
\end{prop}

\begin{prop} [Size\lbrack~P\lbrack $(x,y)$\rbrack~\rbrack]\label{prop23}
{\rm
If Size[ P[$(x,y)$] ] $\le 3^k$ for $k \ge 2$, then 
Size[ left subproblem of P[$(x,y)$] ] $\le 3^{k-1}$, Size[ right subproblem of P[$(x,y)$] ] $\le 3^{k-1}$, and Size[ balanced subproblem of P[$(x,y)$] ] $\le 3^{k-1}$.
}
\end{prop}

\begin{prop} [Problems of size $\le 3$]\label{prop24}
{\rm
Problems P[$(m)$] and P[$(x,y)$] of size $\le 3$ can be solved with at most one weighing.
}
\end{prop}

The maximum number of coins of P[$(m)$] solvable with at most $k$ weighings is $m$ such that $2m=3^k-1$ for $k\ge 1$.  The maximum number of coins of P[$(x,y)$] solvable with at most $k$ weighings is $3^k$, when $x\ne y$ and $x+y=3^k$, or $3^k-1$, when $x=y$ and $x+y=3^k-1$, for $k\ge 1$. 

\section{Related Work}

A Dyson set is a set of vectors where each element is $-1$ or $0$ or $+1$, 
the sum of each dimension part is $0$, and all vectors and their ($-1$) multiplied vectors are distinct. For example, $\lbrace$ ($-1$, $+1$), ($0$, $-1$), ($+1$, $0$) $\rbrace$ is a Dyson set for three-coin problem P[$3$].

\subsection{Dyson method}

Dyson method uses the following Dyson set for the fake coin problem of $4$ coins.  Here $-1$, $0$, and $+1$ respectively stand for the left side, the outside, and the right side for the first and the second weighing. 
Dyson method uses the Dyson set for P[$3$] and the fourth coin $d$ is kept outside for the first and the second weighing.\par\vspace*{5pt}

\begin{tabular}{ll}
$a$: & $(-1, +1)$\\
$b$: & $(0, -1)$\\
$c$: & $(+1, 0)$\\
$d$: & $(0, 0)$\\
\end{tabular}\par\vspace*{5pt}

A vector $v$ of the Dyson set and its
$(-1)$ multiplied vector $-v$ respectively show the results of all weighings
when the coin is heavy and when
the coin is light, where $-1$, $0$, and $+1$
respectively stand for the left side going down,
balanced, and the right side going down.

Based on this Dyson set, the first and the second weighings are as follows. The third weighing is performed when the first and the second weighings are all balanced.

\begin{tabular}{ll}
$1.$ & $a$ versus $c$\\
$2.$ & $b$ versus $a$\\
\multicolumn{2}{l}{(if all balanced)}\\
$3.$ & $d$ versus $a$\\
\end{tabular}

\subsection{Born method}

Born method uses the following Dyson set for $4$ coins.\par\vspace*{5pt}

\begin{tabular}{ll}
$a$: & $(-1, -1, -1)$\\
$b$: & $(+1, 0, 0)$\\
$c$: & $(0, +1, 0)$\\
$d$: & $(0, 0, +1)$\\
\end{tabular}\par\vspace*{5pt}

Based on this Dyson set, three weighings are as follows. All weighings are predetermined.

\begin{tabular}{ll}
$1.$ & $a$ versus $b$\\
$2.$ & $a$ versus $c$\\
$3.$ & $a$ versus $d$\\
\end{tabular}

\subsection{Our Method}

We recursively produce subproblems for a given problem P[$4$]
until we have subproblems of size $\le 3$. We select $9$ as the partition bound for P[$4$].

Bound computation for $4$~~~  $3^{\lceil \log_{3} (2\times4+3)\rceil -1}=3^2=9$

P[$4$]: $4$ $\rightarrow (2,2,0)$ with bound $9$

P[$(2,2)$]: $(2,2)$ $\rightarrow ((1,1),(1,1),(0,0))$ with bound $3$

P[$(0)$], P[$(1,1)$], and P[$(0,0)$] are of size $\le 3$, which can be solved using at most one weighing. 

Based on this solution, when the left side always goes down for example, three weightings are as follows, 

\begin{tabular}{ll}
$1.$ & $a,b$ versus $c,d$ with outside none\\
$2.$ & $a,c$ versus $b,d$ with outside none\\
$3.$ & $a$ versus $c$ (genuine coin) with outside $d$\\
\end{tabular}

\subsection{Problem of 12 coins}

Both Dyson method and Born method use the following Dyson set for P[$12$].

$\lbrace$ $(-1,-1,0)$, $(0,0,+1)$, $(+1,+1,-1)$, $(-1,0,-1)$, $(0,+1,0)$, $(+1,-1,+1)$, $(0,-1,-1)$, $(+1,0,0)$, $(-1,+1,+1)$,
$(-1,0,+1)$, $(0,+1,-1)$, $(+1,-1,0)$ $\rbrace$

Hence the resulting three weighings are as follows.

\begin{tabular}{ll}
$1.$ & $a,d,i,j$ versus $c,f,h,l$ with outside $b,e,g,k$\\
$2.$ & $a,f,g,l$ versus $c,e,i,k$ with outside $b,d,h,j$\\
$3.$ & $c,d,g,k$ versus $b,f,i,j$ with outside $a,e,h,l$\\
\end{tabular}\vspace*{5pt}

Our method produces the following subproblems using partitions with bounds.

Bound computation for $12$~~~$3^{\lceil\log_3 (2\times 12+3)\rceil-1}=3^2=9$\par
P[$12$]: $12$ $\rightarrow (4,4,4)$ with bound $9$\par
P[$(4,4)$]: $(4,4)$ $\rightarrow ((2,1),(2,1),(0,2))$ with bound $3$\par
P[$(2,1)$] and P[$(0,2)$] are of size $\le 3$, which can be solved with one weighing\par
P[$(4)$]: $(4)$ $\rightarrow (2,1,1)$ with bound $3$\par
P[$(2,1)$], P[$(1,2)$], and P[$(1)$] are of size $\le 3$, which can be solved with one weighing.

Based on this solution, when the left side always goes down for example,
three weightings are as follows,

\begin{tabular}{ll}
$1:$ & $a,b,c,d$ versus $e,f,g,h$ with outside $i,j,k,l$\\
$2:$ & $a,b,e$ versus $c,d,f$ with outside $g,h$\\
$3:$ & $a$ versus $b$ with outside $f$\\ 
\end{tabular}

\begin{table}
\caption{P[$(x,y)$] of size $\le3$}
\label{table1}
\begin{tabular}{ccp{0.8\textwidth}}
\hline
Problem & Partition & Description\\
\hline
$(0,0)$ & $-$ & We are done.\\
$(1,0)$ & $-$ & The coin is heavy.\\
$(0,1)$ & $-$ & The coin is light.\\
$(2,0)$ & $(1,0,1)$ & One coin on the left side, another coin on the outside,
 and one genuine coin on the right side.
 If the left side goes down, then it is heavy. 
 If balanced, then the outside coin is heavy.\\
$(1,1)$ & $(1,0,1)$ & One heavy coin candidate on the left side, one light
 coin candidate on the outside, and one genuine coin 
 on the right side.
  If the left side goes down, then it is heavy. 
  If balanced, then the outside coin is light.\\
$(0,2)$ & $(1,0,1)$ & One coin on the left side, another coin on the outside, and one 
genuine coin on the right side.
If the left side goes up, then it is light. 
If balanced, then the outside coin is light.\\
$(3,0)$ & $(1,1,1)$ & One coin on the left side and another clin on the right side.
If one of sides goes down, then the down side coin is heavy. 
If balanced, then the outside coin is heavy.\\
$(2,1)$ & $(1,1,1)$ & One heavy coin candidate on the left side and 
 another heavy coin candidate on the right side.
 If one of sides goes down, then the down side coin is heavy. 
 If balanced, then the outside coin is light.\\
$(1,2)$ & $(1,1,1)$ & One light coin candidate on the left side and 
 another light coin candidate on the right side.
 If one of sides goes up, then the up side coin is light. 
 If balanced, then the outside coin is heavy.\\
$(0,3)$ & $(1,1,1)$ & One coin on the left side and another coin on the right side.
If one of sides goes up, then the up side coin is light. 
If balanced, then the outside coin is light.\\
\hline
\end{tabular}
\end{table}
\begin{table}
\caption{P[$(m)$] of size $\le3$}
\label{table2}
\begin{tabular}{ccp{0.8\textwidth}}
\hline
Problem & Partition & Description\\
\hline
$(0)$ &  $-$  & We are done.\\
$(1)$ & $(1,0,0)$ & One coin on the left side and one genuine coin on the right side.
If the left side goes down, then it is heavy. 
If the left side goes up, then it is light. \\
\hline
\end{tabular}
\end{table}

\section{Induction Basis}

We give solutions to problems P[$(x,y)$] and P[$(m)$] of size $\le 3$.
As shown in Table \ref{table1}, we can confirm that if $x+y\le3$, then P[$(x,y)$] is solvable with at most one weighing.
As shown in Table \ref{table2}, we can confirm that if $2m\le3$, then P[$(m)$] is solvable with at most one weighing. 
Notice that cases P[$(0)$] and P[$(0,0)$] are not selected in actual weighings.

\section{Bounded Partition Method}

We describe our bounded partition method in pseudo-code. 
For producing partitions, we use Standard Partition for P[$n$], Pair Partition for P[$(x,y)$], and Special Partition for P[$(m)$].  To implement P[$n$], P[$(x,y)$], and P[$(m)$], we respectively use procedures F[$c,n$], G[$c,(x,y)$], and H[$c,m$] with partition bound $c$. Partition bound $c$ is computed from given $n$ in procedure FC[$n$]. We use partition bound $3^{k-1}$ for the smallest integer $k$ such that $2n\le 3^k-3$ for given $n$.

In Standard Partition, the number of the left side coins and the number of the right side coins are always same. This number is greater than or equal to the number of the outside coins except when $n=1$. \par

\vspace*{10pt}
StandardPartition[$n$]:\par
If $n=0$ then $(0,0,0)$;\par
If $n=1$ then $(0,0,1)$;\par
If $n=2$ then $(1,1,0)$;\par 
If ($n \ge 3$) and (Mod[$n,3$]$=0$) then $\left( \frac{n}{3}, \frac{n}{3}, \frac{n}{3}\right)$;\par\hspace*{3ex}  $\lbrace (j,j,j)$ if $n=3j\rbrace$\par
If ($n \ge 3$) and (Mod[$n,3$]$=1$) then $\left( \frac{n+2}{3}, \frac{n+2}{3}, \frac{n-4}{3}\right)$;\par\hspace*{3ex} $\lbrace (j+1,j+1,j-1)$ if $n=3j+1\rbrace$\par
If ($n \ge 3$) and (Mod[$n,3$]$=2$) then $\left( \frac{n+1}{3}, \frac{n+1}{3}, \frac{n-2}{3}\right)$;\par\hspace*{3ex} $\lbrace (j+1,j+1,j)$ if $n=3j+2\rbrace$
\vspace*{10pt}

In Special Partition, the number of the left side coins and the number of the right side coins are not same for $m=3j+1$ $(j\ge 0)$ case or $m=2$ case. For these weighing cases, we add one genuine coin on the right side so that we have the same number of coins on both sides.

\vspace*{10pt}
SpecialPartition[$m$]:\par
If $m=0$ then $(0,0,0)$;\par
If $m=1$ then $(1,0,0)$;\par
If $m=2$ then $(1,0,1)$;\par 
If ($m \ge 3$) and (Mod[$m,3$]$=0$) then $\left( \frac{m}{3}, \frac{m}{3}, \frac{m}{3} \right)$;\par\hspace*{3ex}  $\lbrace (j,j,j)$ if $m=3j \rbrace$\par
If ($m \ge 3$) and (Mod[$m,3$]$=1$) then $\left( \frac{m+2}{3}, \frac{m-1}{3}, \frac{m-1}{3} \right)$;\par\hspace*{3ex}  $\lbrace (j+1,j,j)$ if $m=3j+1\rbrace$\par
If ($m \ge 3$) and (Mod[$m,3$]$=2$) then $\left(\frac{m+1}{3}, \frac{m+1}{3}, \frac{m-2}{3}\right)$;\par\hspace*{3ex}  $\lbrace (j+1,j+1,j)$ if $m=3j+2\rbrace$\par
\vspace*{10pt}

Pair Partition produces $((x_1,y_1),(x_2,y_2),(x_3,y_3))$ such that $x_1=x_2, y_1=y_2, x_1+y_1\le c$ and $x_2+y_2\le c$ for $(x, y)$ such that $x+y>3$. When $x_1 + y_1 > c$ or $x_2 + y_2 > c$, overflow coins are added to $y_3$.

\vspace*{10pt}
PairPartition[$c, (x,y)$]:\par
If $x+y\le 3$ then $((x, y), (0, 0), (0, 0))$ else (\par
\hspace*{10pt}$(x_1, x_2, x_3)$ $\leftarrow$ StandardPartition[$x$];\par
\hspace*{10pt}$(y_1, y_2, y_3)$ $\leftarrow$ StandardPartition[$y$];\par
\hspace*{10pt}If $x_1+y_1 > c$ then $(y_1, y_3) \leftarrow (c-x_1,  y_3+y_1+x_1-c)$;\par
\hspace*{10pt}If $x_2+y_2 > c$ then $(y_2, y_3)\leftarrow (c-x_2, y_3+y_2+x_2-c)$;\par
\hspace*{10pt}$((x_1,y_1),(x_2,y_2),(x_3,y_3))$;\par
);\par
\vspace*{10pt}

\vspace*{10pt}
FC[$n$]:\par
$c \leftarrow 3^{\lceil \log_{3} (2n+3)\rceil -1}$; F[$c,n$];
\vspace*{10pt}

As for availability of the use of one genuine coin, notice that if $n \ge 3$ and $n \ne 4$, then we can use one genuine coin after the first weighing by Standard Partition.
If $n = 4$, then we can use one genuine coin after the second weighing by Pair Partition. 

\vspace*{10pt}
F[$c, n$]:\par
$(x, y, m) \leftarrow$ StandardPartition[$n$];\par
Print[$c,(x,y,m)$];\par 
If the left side goes down and $x+y>3$ then G[$\frac{c}{3}, (x, y)$];\par
If the right side goes down and $y+x>3$ then G[$\frac{c}{3}, (y, x)$];\par
If balanced and $2m>3$ then H[$\frac{c}{3}, m$];
\vspace*{10pt}

G[$c, (x,y)$]:\par
$((x_1,y_1), (x_2,y_2), (x_3,y_3)) \leftarrow$ PairPartition[$c, (x, y)$];\par
Print[$c, ((x_1,y_1), (x_2,y_2), (x_3,y_3))$];\par 
If the left side goes down and $x_1+y_2>3$ then G[$\frac{c}{3}, (x_1, y_2)$];\par
If the right side goes down and $x_2+y_1>3$ then G[$\frac{c}{3}, (x_2, y_1)$];\par
If balanced and $x_3+y_3>3$ then G[$\frac{c}{3}, (x_3, y_3)$];\par

\vspace*{10pt}
H[$c, m$]:\par
$(x,y, t) \leftarrow$ SpecialPartition[$m$];\par
Print[$c,(x,y,t)$];\par\hspace*{3ex}
\begin{minipage}{0.9\textwidth}$\lbrace$If $y=x-1$ then we add one genuine coin on the right side for weighing purpose only$\rbrace$\end{minipage}\par
If the left side goes down and $x+y>3$ then G[$\frac{c}{3}, (x, y)$];\par
If the right side goes down and $y+x>3$ then G[$\frac{c}{3}, (y, x)$];\par
If balanced and $2t>3$ then H[$\frac{c}{3}, t$];\par
\vspace*{10pt}

Notice that Special Partition and Pair Partition are well-behaved partitions. 
But Standard Partition is not well-behaved. If we use Standard Partition for P[$(4)$], then Standard Partition of $4$ is $(2, 2, 0)$. It is not acceptable with bound $3$, where $2\times 4\le 9$ and $2+2 > 3$.  Special Partition of $4$ is $(2, 1, 1)$ and it is acceptable with bound $3$, because $2\times 4\le 9$, $2+1=1+2\le 3$ and $2\times 1\le 3$. 

\begin{exam*}[{Problem of 39 coins}]\label{ex2}
{\rm
Subproblems of $4$ weighings for P[$39$] are as follows.\par
Bound computation for $39$~~~$3^{\lceil\log_3 (2\times 39+3)\rceil-1}=3^3=27$\par
P[$39$]: $39$ $\rightarrow (13,13,13)$ with bound $27$\par
P[$(13,13)$]: $(13,13)$ $\rightarrow ((5,4),(5,4),(3,5))$ with bound $9$\par
P[$(5,4)$]: $(5,4)$ $\rightarrow ((2,1),(2,1),(1,2))$ with bound $3$\par
P[$(3,5)$]: $(3,5)$ $\rightarrow ((1,2),(1,2),(1,1))$ with bound $3$\par
P[$(13)$]: $(13)$ $\rightarrow (5,4,4)$ with bound $9$\par
P[$(5,4)$]: $(5,4)$ $\rightarrow ((2,1),(2,1),(1,2))$ with bound $3$\par
P[$(4, 5)$]: $(4, 5)$ $\rightarrow ((2, 1), (2, 1), (0, 3))$ with bound $3$\par
P[$(4)$]: $(4)$ $\rightarrow (2,1,1)$ with bound $3$\par
P[$(2,1)$], P[$(1,2)$], P[$(1,1)$], P[$(0, 3)$], and P[$(1)$] are of size $\le 3$, which can be solved with one weighing.
}
\end{exam*}

These procedures, F[$c, n$], G[$c,(x, y)$], and H[$c, m$], replace
one problem by its subproblem recursively
until the size of a subproblem $\le 3$ in search mode.
If we delete component conditions, "the left side goes down and",
"the right side goes down and", and
"balanced and", from if-conditions of these procedures, then we
can enumerate all subproblems until the size of subproblems $\le 3$.

\section{Concluding Remarks}

We have shown our bounded partition approach to
the fake coin problems. With the help of one genuine coin available
after the first or the second weighing, we can naturally have the three subproblems
of a given problem.  Our method is essentially a direct ternary search method for finding one fake coin based on the size of subproblems solvable with fewer weighings than the given problem.


\bibliographystyle{elsarticle-num}
\bibliography{refs}

\begin{thebibliography}{1}
\expandafter\ifx\csname url\endcsname\relax
  \def\url#1{\texttt{#1}}\fi
\expandafter\ifx\csname urlprefix\endcsname\relax\def\urlprefix{URL }\fi
\expandafter\ifx\csname href\endcsname\relax
  \def\href#1#2{#2} \def\path#1{#1}\fi

\bibitem{Dyson}
F.~J. Dyson, The problem of the pennies, The Mathematical Gazette 30~(291)
  (1946) 231--234.

\bibitem{Born}
A.~Born, C.~A.~J. Hurkens, G.~J. Woeginger, How to detect a counterfeit coin:
  Adaptive versus non-adaptive solutions, Information Processing Letters 86~(3)
  (2003) 137--141.

\end{thebibliography}

\section*{Appendix}

\setcounter{section}{2}
\begin{prop} [F\lbrack $c, n$\rbrack : $(x,y, m) \leftarrow$ StandardPartition\lbrack $n$\rbrack]\label{a-prop21}
{\rm
If $6\le 2n\le 3^k-3$ for $k\ge 2$, then $x+y\le 3^{k-1}, y+x\le 3^{k-1}$, and 
$2m\le 3^{k-1}$.
}
\end{prop}

\newproof{pot21}{Proof of Property \ref{a-prop21}}
\newproof{pot22}{Proof of Property \ref{a-prop22}}
\newproof{pot23}{Proof of Property \ref{a-prop23}}

\begin{pot21}
~\par
Case $1$. 
$n=3j$ and $(x,y, m)=(j, j, j)$ for $j\ge 1$\par
If $2n=6j\le 3^k-3$, then $2j\le 3^{k-1}-1\le 3^{k-1}$.\par
Hence $x+y=y+x\le 3^{k-1}$ and $2m\le 3^{k-1}$.\par\vspace*{5pt}
Case $2$. 
$n=3j+1$ and $(x,y, m)=(j+1, j+1, j-1)$ for $j\ge 1$\par 
If $2n=6j+2 \le 3^k-3$, then $6j+6 \le 3^k+1$.\par
Hence $2m=2j-2$, $x+y=y+x=2j+2$, and $2j-2<2j+2 \le 3^{k-1}+\frac{1}{3}$.\par 
Therefore $2j+2 \le 3^{k-1}$, because both $2j+2$ and $3^{k-1}$ are integers.
Notice that if we change the required precondition from
$2n\le 3^k-3$ to $2n\le 3^k$, then $2j+2\le 3^{k-1}+\frac{4}{3}$ and $x+y$ can be larger than $3^{k-1}$. 
\par\vspace*{5pt}
Case $3$. 
$n=3j+2$ and $(x,y, m)=(j+1, j+1, j)$ for $j\ge 1$\par 
If $2n=6j+4\le 3^k-3$, then $6j+6\le 3^k-1$.\par
Hence $2m=2j$, $x+y=y+x=2j+2$, and $2j<2j+2\le 3^{k-1}-\frac{1}{3}\le 3^{k-1}$.
\end{pot21}

\begin{prop}[H\lbrack $c, m$\rbrack : $(x,y,t) \leftarrow$ SpecialPartition\lbrack $m$\rbrack]\label{a-prop22}
{\rm
If $2m\le 3^k$ for $k\ge 2$, then $x+y\le 3^{k-1}, y+x\le 3^{k-1}$, and 
$2t\le 3^{k-1}$.
}
\end{prop}

\begin{pot22}
~\par
When $m=2$ and $(2) \rightarrow (1,0,1)$, this property holds. Cases where $m\ge 3$ are as follows.\par
Case $1$. 
$m=3j$ and $(x, y, t)=(j, j, j)$ for $j\ge 1$\par 
If $2m=6j\le 3^k$, then $2j\le 3^{k-1}$.\par
Hence $x+y=y+x=2j \le 3^{k-1}$ and $2t=2j \le 3^{k-1}$.
\par\vspace*{5pt}
Case $2$. 
$m=3j+1$ and $(x, y, t)=(j+1, j, j)$ for $j\ge 1$\par 
If $2m=6j+2\le 3^k$, then $6j+3\le 3^k+1$.\par
Hence $2t=2j$, $x+y=y+x=2j+1$, and $2j<2j+1\le 3^{k-1}+\frac{1}{3}$.\par
Therefore $2j+1\le 3^{k-1}$, because both $2j+1$ and $3^{k-1}$ are integers.\par\vspace*{5pt}
Case $3$. 
$m=3j+2$ and $(x, y, t)=(j+1, j+1, j)$ for $j\ge 1$\par 
If $2m=6j+4\le 3^k$, then $6j+6\le 3^k+2$.\par
Hence $2t=2j$, $x+y=y+x=2j+2$, $2j<2j+2\le 3^{k-1}+\frac{2}{3}$.\par 
Therefore $2j+2\le 3^{k-1}$, because both $2j+2$ and $3^{k-1}$ are integers.
\end{pot22}

\begin{prop}[G\lbrack $c, (x,y)$\rbrack : $((x_1,z_1), (x_2, z_2), (x_3, z_3)) \leftarrow$ PairPartition\lbrack $c, (x, y)$\rbrack]\label{a-prop23}
{\rm
$(x_1, x_2, x_3) \leftarrow$ StandardPartition[$x$];
$(y_1, y_2, y_3) \leftarrow$ StandardPartition[$y$]; and
Pair Partition with bound $c$ produces $((x_1,z_1),(x_2,z_2),(x_3,z_3))$.
If $x+y\le 3^k$ for $k\ge 2$, then $c=3^{k-1}$, $x_1+z_2\le 3^{k-1}$, $x_2+z_1\le 3^{k-1}$,
and $x_3+z_3\le 3^{k-1}$.
}
\end{prop}

\begin{pot23}
~\par General Case: When neither $x$ nor $y$ is $1$.\par
Because $x_1=x_2$, $y_1=y_2$, and $z_1=z_2$, it suffices to show that if
$x+y\le 3^k$ for $k\ge 2$, then $x_1+z_1\le 3^{k-1}$, $x_2+z_2\le 3^{k-1}$,
and $x_3+z_3\le 3^{k-1}$.
Notice that $c=3^{k-1}$, $x+y\le 3c=3^k$, $x=x_1+x_2+x_3$, and $y=y_1+y_2+y_3$.
\par
Case $1$.\par
If $x_1+y_1\le c$ and $x_2+y_2\le c$, then $(z_1,z_2,z_3)=(y_1,y_2,y_3)$.\par
Hence $x_3\le x_1=x_2$, $y_3\le y_1=y_2$ and $x_3+y_3\le x_1+y_1\le c$.
\par\vspace*{5pt}
Case $2$.\par
If $x_1+y_1>c$ and $x_2+y_2>c$, then
$(x_1+z_1, x_2+z_2, x_3+z_3)=(c, c, x_3+y_3+(x_1+y_1-c)+(x_2+y_2-c))$.\par
If $x+y\le 3c$, then $x+y=x_1+x_2+x_3+z_1+z_2+z_3\le 3c$.\par
Hence $x_3+z_3\le 3c-(x_1+z_1+x_2+z_2)=3c-2c=c$.
\par\vspace*{5pt}
Exceptional Case: When $x$ or $y$ is $1$.\par
P[$(x,y)$] can be P[$(1,1)$] or P[$(x,1)$] or P[$(1,y)$] for $x\ne 1$ and $y\ne 1$.
It suffices to show that three cases of P[$(x,1)$],
cases of $x+1=3^k$, $x+1=3^k-1$, and $x+1=3^k-2$ for $k\ge 2$,
produce the size of subproblems $\le 3^{k-1}$.\par
Case 1.\par
If $x=3^k-1=3(3^{k-1}-1)+2$, then
$(x,1) \rightarrow ((3^{k-1}-1+1,0),(3^{k-1}-1+1,0),(3^{k-1}-1,1))$.\par
Case 2.\par
If $x=3^k-2=3(3^{k-1}-1)+1$, then
$(x,1) \rightarrow ((3^{k-1}-1+1,0),(3^{k-1}-1+1,0),(3^{k-1}-1-1,1))$.\par
Case 3.\par
If $x=3^k-3=3(3^{k-1}-1)$, then
$(x,1) \rightarrow ((3^{k-1}-1,0),(3^{k-1}-1,0),(3^{k-1}-1,1))$.\par
Hence the size of all subproblems $\le 3^{k-1}$. 
\end{pot23}

\end{document}